\documentclass[a4paper,5p,fleqn]{cas-dc}
\usepackage[numbers]{natbib}
\usepackage{multirow}
\usepackage{siunitx}
\usepackage{gensymb}
\usepackage[table]{xcolor}
\definecolor{lightgreen}{RGB}{144, 238, 144}
\definecolor{lightorange}{RGB}{255, 200, 100}
\definecolor{lightred}{RGB}{255, 100, 100}

\def\tsc#1{\csdef{#1}{\textsc{\lowercase{#1}}\xspace}}
\tsc{WGM}
\tsc{QE}
\tsc{EP}
\tsc{PMS}
\tsc{BEC}
\tsc{DE}

\begin{document}
\let\WriteBookmarks\relax
\def\floatpagepagefraction{1}
\def\textpagefraction{.001}
\shorttitle{Automated Histopathologic Assessment of Hirschsprung Disease Using a Multi-Stage Vision Transformer Framework}
\shortauthors{Y. Megahed and S. Abou-Alwan et~al.}

\title [mode = title]{Automated Histopathologic Assessment of Hirschsprung Disease Using a Multi-Stage Vision Transformer Framework}                      



\author[1,4]{Youssef Megahed}[type=editor,
                        orcid=0009-0004-2595-5468]                  
\cormark[1]
\ead{youssefmegahed@cmail.carleton.ca}

\credit{Conceptualization of this study, Methodology, Software}

\affiliation[1]{organization={Department of Systems and Computer Engineering, Carleton University}, 
                city={Ottawa},
                state={Ontario},
                country={Canada}}
                
\affiliation[2]{organization={Department of Clinical Science and Translational Medicine, University of Ottawa}, 
                city={Ottawa},
                state={Ontario},
                country={Canada}}

\affiliation[3]{organization={School of Epidemiology and Public Health, University of Ottawa}, 
                city={Ottawa},
                state={Ontario},
                country={Canada}}

\affiliation[4]{organization={Department of Methodological and Implementation Research, Ottawa Hospital Research Institute}, 
                city={Ottawa},
                state={Ontario},
                country={Canada}}

\affiliation[5]{organization={Children’s Hospital of Eastern Ontario (CHEO)}, 
                city={Ottawa},
                state={Ontario},
                country={Canada}}

\author[1]{Saleh Abou-Alwan}
\cormark[1]
\author[1]{Anthony Fuller}
\author[5]{Dina El Demellawy}
\author[1,2,3,4]{Steven Hawken}
\cormark[2]
\author[1]{Adrian D. C. Chan}
\cormark[2]
\ead{adrianchan@cunet.carleton.ca}


\credit{Data curation, Writing - Original draft preparation}


\cortext[cor1]{Co-first author}
\cortext[cor2]{Co-advising author}

\begin{abstract}
Hirschsprung Disease is characterized by the absence of ganglion cells in the myenteric plexus. Therefore, the correct identification of ganglion cells is crucial for diagnosing Hirschsprung disease. We introduce a three-stage analysis framework that mimics the pathologist's diagnostic approach. The framework, based on a Vision Transformer model (ViT-B/16), sequentially segments the muscularis propria, segments the myenteric plexus, and detects ganglion cells within anatomically valid regions. 30 whole-slide images of colon tissue were used, each containing manual annotations of muscularis, plexus, and ganglion cells. A 5-fold cross-validation scheme was applied to each stage, along with resolution-specific tiling strategies and tailored postprocessing to ensure anatomical consistency. The proposed method achieved a Dice coefficient of 89.9\% and a Plexus Inclusion Rate of 100\% for muscularis segmentation. Plexus segmentation reached a recall of 94.8\%, a precision of 84.2\% and a Ganglia Inclusion Rate of 99.7\%. For ganglion cells annotated with high certainty, the model achieved 62.1\% precision and 89.1\% recall. When considering all annotated ganglion cells, regardless of certainty level, the overall precision was 67.0\%. These results indicate that ViT-based models are effective at leveraging global tissue context and capturing cellular morphology at small scales, even within complex histological tissue structures. This multi-stage methodology has great potential to support digital pathology workflows by reducing inter-observer variability and assisting in the evaluation of Hirschsprung disease. The clinical impact will be evaluated in future work with larger multi-center datasets and additional expert annotations.
\end{abstract}




\maketitle

\section{Introduction}
Hirschsprung disease (HD), a congenital birth defect defined by the absence of ganglionic cells within the myenteric plexus of the intestinal tract {\hypersetup{hidelinks}\textcolor{blue}{\cite{b7}}}, affects 1/5000 neonatal patients worldwide {\hypersetup{hidelinks}\textcolor{blue}{\cite{b1}}}. Absence of ganglionic cells is due to defective migration, proliferation and differentiation of neural crest cells in the womb. It leads to malformation of enteric nerve cells that cause obstruction of the bowel. The patient presents with symptoms of severe constipation and features of intestinal obstruction {\hypersetup{hidelinks}\textcolor{blue}{\cite{b2}}}. Thus, it is necessary to identify the ganglion cells of the colon to confirm HD.

After the onset of symptoms, traditional methods for diagnosing HD include a contrast enema radiograph, in which a barium contrast is inserted into the child's rectum and x-ray images of the colon are taken to visualize the structure of different colon segments, thereby confirming the presence of malformed intestines. A second common method for diagnosing HD is a rectal biopsy, in which a section of tissue is extracted from the rectum and examined on a Whole Slide Image (WSI) by a trained pathologist. This method involves quantitatively assessing the presence of ganglion cells. Although both approaches are considered the gold standard for confirming HD, they have significant limitations, including sampling error, inter- and intra-rater variability among pathologists {\hypersetup{hidelinks}\textcolor{blue}{\cite{b8,b10}}}, and substantial time requirements. More critically, rectal biopsies are used to determine the most distal healthy segment of the intestine that can be preserved during a surgical pull-through procedure, which is a primary treatment option. Misidentifying a healthy region as aganglionic may lead to unnecessary resection of functional bowel, while misclassifying an aganglionic segment as healthy can result in surgical failure, which is particularly consequential given the limited physiological reserve of newborn patients, highlighting the importance of automating the identification of ganglionic cells with high accuracy. 

Recent advances in computational imaging and digital pathology have enabled Artificial Intelligence (AI) methods to automate the classification {\hypersetup{hidelinks}\textcolor{blue}{\cite{b31}}}, detection {\hypersetup{hidelinks}\textcolor{blue}{\cite{b32}}}, and segmentation {\hypersetup{hidelinks}\textcolor{blue}{\cite{b33}}} of histopathology images across a wide range of organ systems. For instance, several Deep Learning (DL) algorithms have been utilized for computer-aided diagnosis, attempting to classify breast cancer lesions using WSIs{~\hypersetup{hidelinks}\textcolor{blue}{\cite{b3}}} and human aortic and cardiac regions {\hypersetup{hidelinks}\textcolor{blue}{\cite{b5}}}. Other algorithms{~\hypersetup{hidelinks}\textcolor{blue}{\cite{b4}}} have been deployed to segment images of bronchoscopes, further identifying lung-related diseases. Furthermore, object segmentation has also been widely applied to various medical images and diseases. Specifically, classifying the potential onset of autism spectrum disorder (ASD), including additional indicators of ASD and irregular brain structure {\hypersetup{hidelinks}\textcolor{blue}{\cite{b24,b25,b30}}}. Leveraging DL based computer vision for identifying ganglion cells is a promising direction. An effective strategy is to incorporate anatomical context into the workflow, since ganglion cells reside within the myenteric plexus, which itself is embedded in the muscularis layer. This motivates a multistage approach in which the muscularis is first identified, the plexus is then localized within that region, and ganglion cells are subsequently detected within the predicted plexus to confirm the presence of HD.

\begin{figure*}
	\centering
	\includegraphics[width=0.7\textwidth]{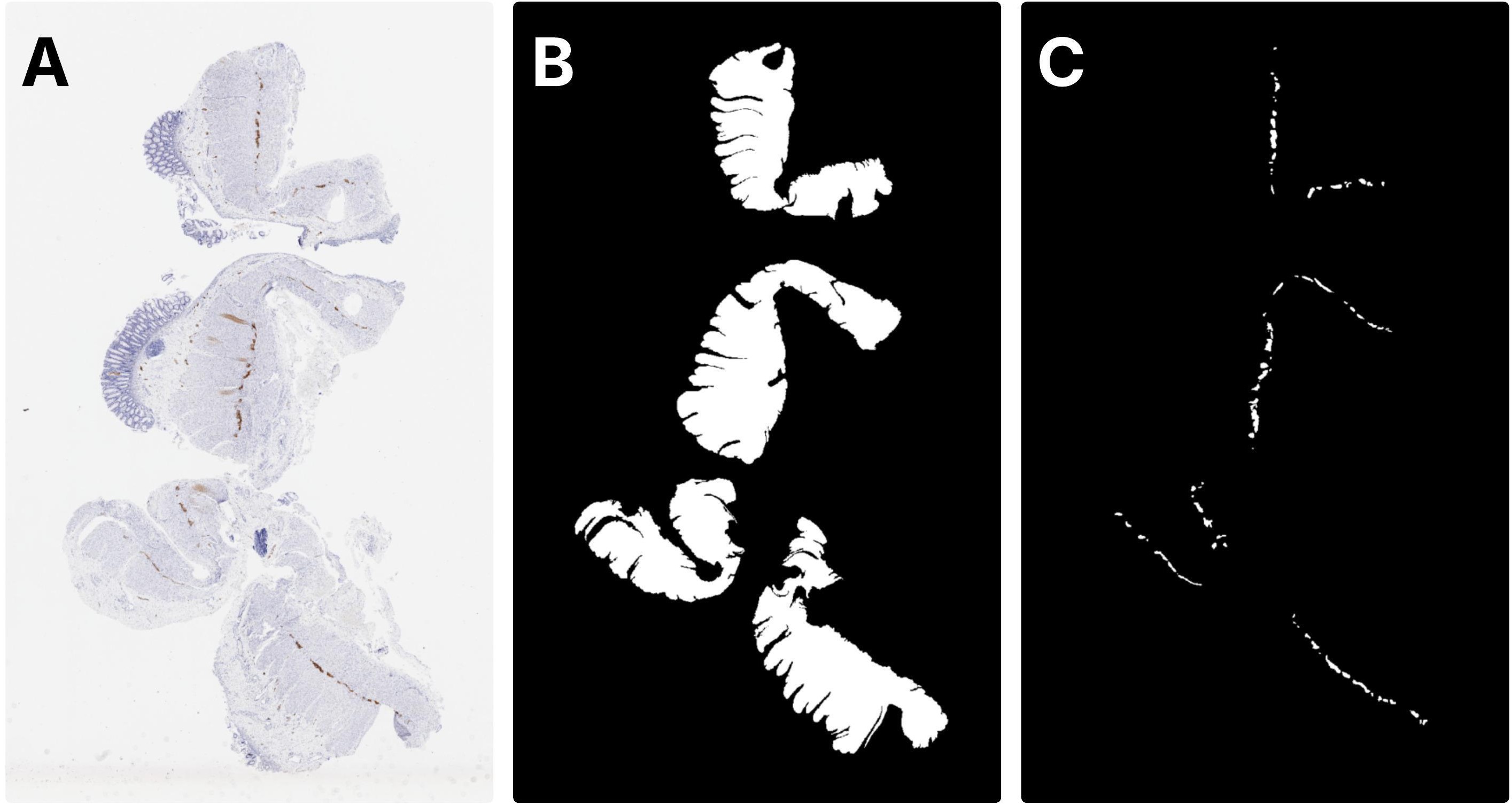}
	\caption{Reference annotations for tissue compartments in a whole slide section.
    \textbf{A:} Original WSI.
    \textbf{B:} Ground truth mask delineating the muscularis propria layer.
    \textbf{C:} Ground truth mask marking the myenteric plexus regions.}
	\label{FIG:muscle_example}
\end{figure*}

Recent attempts{~\hypersetup{hidelinks}\textcolor{blue}{\cite{b11,b12,b13,b14,b15}}} have been made to segment the muscularis layer, which encloses the myenteric plexus, first utilizing a shallow machine learning base model, K-means clustering{~\hypersetup{hidelinks}\textcolor{blue}{\cite{b14}}}. This was then followed by deploying a Convolutional Neural Network (CNN), which produced a far greater accuracy and plexus inclusion rate when segmenting the muscularis layer{~\hypersetup{hidelinks}\textcolor{blue}{\cite{b13,b14}}}. A colour-thresholding algorithm was then utilized for segmenting the myenteric plexus region and effectively detecting Carletinin-positive ganglia inside plexus regions. Finally, ganglia were segmented from each plexus region using intensity thresholding on the previously segmented Calretinin signal. This was followed by deploying a Linear Discriminant Analysis (LDA) model to identify true-ganglia from false-ganglia{~\hypersetup{hidelinks}\textcolor{blue}{\cite{b14}}}. Overall, the results suggest strong discriminative ability, offering a promising avenue for automated DL. Other attempts have been made {\hypersetup{hidelinks}\textcolor{blue}{\cite{b6}}} that leverage the U-Net, a CNN algorithm, to similarly detect the presence of ganglion cells in the myenteric plexus, using WSIs from biopsy tissues. M. Duci et. al., {\hypersetup{hidelinks}\textcolor{blue}{\cite{b6}}} demonstrated innate capabilities of the CNN architecture, especially given its ability to extract low-level and high-level features throughout different layers, in addition to capturing local relationships through convolutions. Overall, the algorithm is able to learn rich, essential semantic features from the colon layers in the WSI dataset. Although U-Net is appropriate for learning pixel-wise local representations, it cannot learn long-range, global contextual relationships across different regions of the image, which could be critical for detecting ganglion cells dispersed throughout the WSI. 

Transformers were first introduced in Natural Language Processing (NLP), where their attention-based architecture demonstrated a strong ability to model long-range dependencies. Their success led to rapid adoption in computer vision, where Vision Transformers (ViTs) {\hypersetup{hidelinks}\textcolor{blue}{\cite{b17}}} have shown competitive and often superior performance to CNNs across general imaging benchmarks {\hypersetup{hidelinks}\textcolor{blue}{\cite{b17,b35,b36}}} and medical image analysis tasks {\hypersetup{hidelinks}\textcolor{blue}{\cite{b9,b15,b16,b30}}}. This is possible because ViTs have inherent self-attention {\hypersetup{hidelinks}\textcolor{blue}{\cite{b34}}}, which enables them to model long-range dependencies and capture global contextual relationships across all parts of an image. Thereby, recognizing patterns that span large areas, such as tissue structure in WSI histopathology. 

In this study, we introduce a three-stage semantic segmentation approach that mirrors the colon's anatomical hierarchy. The first stage identifies the muscularis layer, which provides the outer boundary within which the myenteric plexus regions are located. The second stage segments the myenteric plexus regions inside the predicted muscularis. The third stage then detects ganglion cells "objects" within the plexus region, enabling assessment of ganglion cell presence for HD diagnosis. The study evaluates a three-stage ViT-based pipeline that processes WSIs in a tile-based manner. In each stage, the WSI resolution is adjusted so that a 224$\times$224 tile captures an appropriate field of view (FOV) for the anatomical structure of interest while still preserving the level of detail needed for accurate segmentation. This allows the model to access broad contextual information for larger structures, such as the muscularis, and finer morphological features for smaller structures, such as ganglion cells. 

Performance is assessed at each stage of the pipeline and for the combined workflow, and results are compared with previously published methods that used the same dataset {\hypersetup{hidelinks}\textcolor{blue}{\cite{b11,b13,b14}}}. This framing highlights the contribution of each stage to the overall identification task and enables a direct comparison with earlier approaches that relied on shallow learning or CNN-based models.

\begin{figure*}
	\centering
	\includegraphics[width=0.95\textwidth]{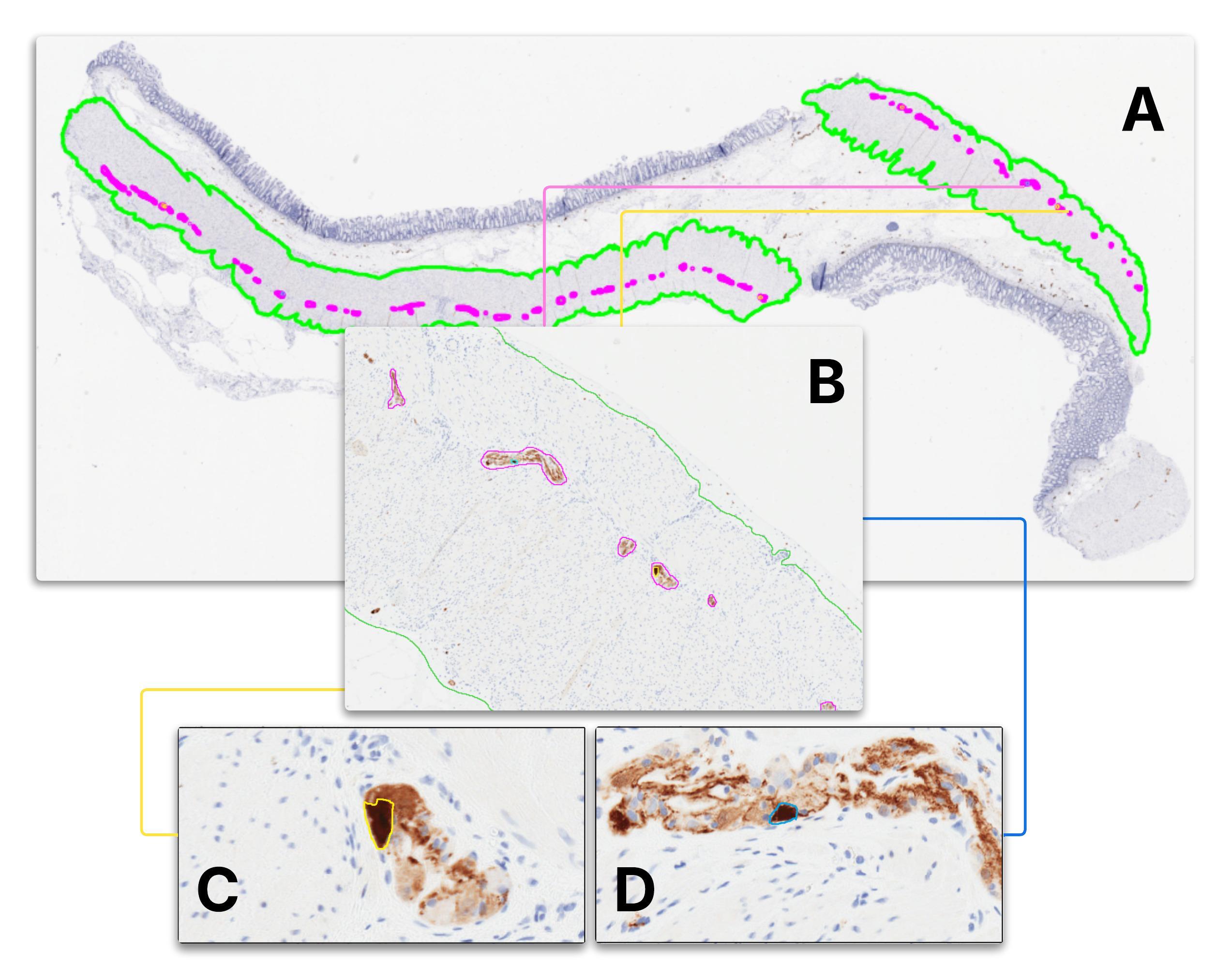}
	\caption{Overview of tissue regions on a whole slide section.
    \textbf{A:} Low magnification view with the muscularis Propria tissue boundary outlined in green and the myenteric plexus path marked in magenta. 
    \textbf{B:} Intermediate magnification showing myenteric plexus regions highlighted in magenta within the annotated muscularis propria segment.
    \textbf{C:} Close view of a ganglion cell (within the myenteric plexus regions) with high certainty outlined in yellow.
    \textbf{D:} Close view of a ganglion cell (within the myenteric plexus regions) with low certainty in cyan.}
	\label{FIG:stages_visualization}
\end{figure*}

\section{Methodology}
\subsection{Dataset}
The dataset used in this research study originates from the Children's Hospital of Eastern Ontario (CHEO), comprising 30 WSIs corresponding to colon sections from 26 different patients, all of whom were diagnosed with HD. All high-resolution images were extracted from prepared tissue slides after being scanned using the Aperio ScanScope CS (Aperio Technologies) at a 20$\times$ magnification level (0.50 \si{\micro\metre}/pixel). All 30 WSIs were saved in an SVS format. For each layer, there exists ground truth annotations that a graduate researcher manually segmented (and approved by a pediatric pathologist) {\hypersetup{hidelinks}\textcolor{blue}{\cite{b14}}} ({\hypersetup{hidelinks}\textcolor{blue}{Fig.~\ref{FIG:muscle_example}}}). For instance, the muscularis propria was manually delineated to represent the ground-truth segmentation ({\hypersetup{hidelinks}\textcolor{blue}{Fig.~\ref{FIG:stages_visualization}A}}), the myenteric plexus region was roughly manually segmented (a visually noticeable amount of tissue around the plexus regions was also included) ({\hypersetup{hidelinks}\textcolor{blue}{Fig.~\ref{FIG:stages_visualization}B}}). The ganglion cells were also roughly segmented with a confidence level accompanies each ganglion cell annotation. A high-confidence level indicates strong certainty that the annotated object is a ganglion cell ({\hypersetup{hidelinks}\textcolor{blue}{Fig.~\ref{FIG:stages_visualization}C}}). In contrast, a low-confidence level reflects that the object is believed to be a ganglion cell but with some uncertainty ({\hypersetup{hidelinks}\textcolor{blue}{Fig.~\ref{FIG:stages_visualization}D}}), due to the difficulty in recognizing and selecting physiological features of a ganglion cell {\hypersetup{hidelinks}\textcolor{blue}{\cite{b15}}}.

\subsection{Data Preprocessing}
\subsubsection{Muscularis Propria}
Preprocessing for the muscularis propria region followed a three-step procedure. First, WSIs were downsampled from the original 20$\times$ (0.50 \si{\micro\metre}/pixel) resolution to 5 \si{\micro\metre}/pixel (equivalent to a 10$\times$ reduction in spatial resolution). At this lower resolution, each 224$\times$224 tile covers a larger physical FOV, which is appropriate for capturing the broad, continuous architecture of the muscularis propria. Second, colour normalization was applied using the Macenko method{~\hypersetup{hidelinks}\textcolor{blue}{\cite{b18}}} to mitigate staining variation across slides. Third, each WSI was tiled into 224$\times$224-pixel subimages, from which 1000 tiles were randomly selected per slide (total of 30,000 tiles). These tiles enabled the model to learn characteristic muscularis patterns across the dataset. An example of the corresponding ground truth annotations is shown in{~\hypersetup{hidelinks}\textcolor{blue}{Fig.~\ref{FIG:muscle_example}B}}.

\begin{figure*}
	\centering
	\includegraphics[width=0.7\textwidth]{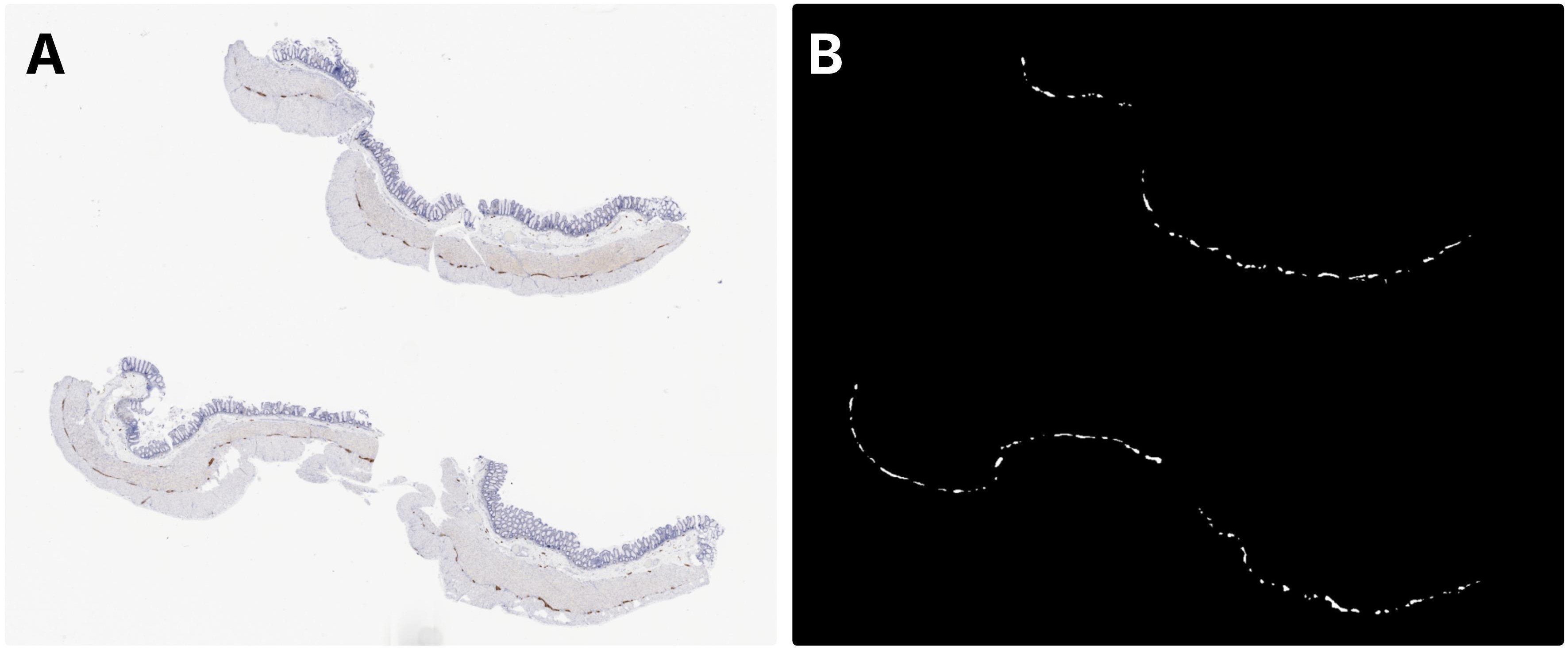}
	\caption{Reference annotations for tissue compartments in a whole slide section.
    \textbf{A:} Original WSI.
    \textbf{B:} Ground truth mask marking the myenteric plexus regions.}
	\label{FIG:plexus_example}
\end{figure*}

\subsubsection{Myenteric Plexus}
Preprocessing for the myenteric plexus followed the same general workflow but used a resolution tailored to plexus morphology. WSIs were downsampled from the original 20$\times$ (0.50 \si{\micro\metre}/pixel) resolution to 2.5 \si{\micro\metre}/pixel (equivalent to a 5$\times$ reduction in spatial resolution). At this level, each 224$\times$224 tile spans a FOV wide enough to capture elongated plexus structures while still preserving sufficient detail to distinguish them from surrounding muscularis tissue. Macenko normalization{~\hypersetup{hidelinks}\textcolor{blue}{\cite{b18}}} was applied for stain harmonization. WSIs were then tiled into 224$\times$224 pixel patches, and tiles were retained only if they contained at least one pixel from the muscularis propria region. Since the myenteric plexus is anatomically confined to the muscularis, this ensured that only relevant tissue regions were included. Examples of plexus annotations are shown in{~\hypersetup{hidelinks}\textcolor{blue}{Fig.~\ref{FIG:muscle_example}C}} and{~\hypersetup{hidelinks}\textcolor{blue}{Fig.~\ref{FIG:plexus_example}B}}.

\subsubsection{Ganglion Cells}
Preprocessing for ganglion cell analysis also followed this framework, but used a higher-resolution image appropriate for cellular-level features. WSIs were downsampled from the original 20$\times$ (0.50 \si{\micro\metre}/pixel) resolution to 1.0 \si{\micro\metre}/pixel (equivalent to a 2$\times$ reduction in spatial resolution). This higher resolution allows each 224$\times$224 tile to preserve fine morphological details necessary for identifying ganglion cells. Macenko normalization{~\hypersetup{hidelinks}\textcolor{blue}{\cite{b18}}} was again applied. WSIs were tiled into 224$\times$224 pixel patches, and tiles were retained only if they contained at least one pixel from the myenteric plexus regions. Because ganglion cells occur exclusively within the plexus, this ensured that model training was restricted to anatomically relevant regions. Examples of ganglion cells are shown in{~\hypersetup{hidelinks}\textcolor{blue}{Fig.~\ref{FIG:stages_visualization}C}} and{~\hypersetup{hidelinks}\textcolor{blue}{Fig.~\ref{FIG:stages_visualization}D}}.

\subsection{Training \& Testing}
\subsubsection{Model Architecture}
\begin{figure*}
	\centering
	\includegraphics[width=0.98\textwidth]{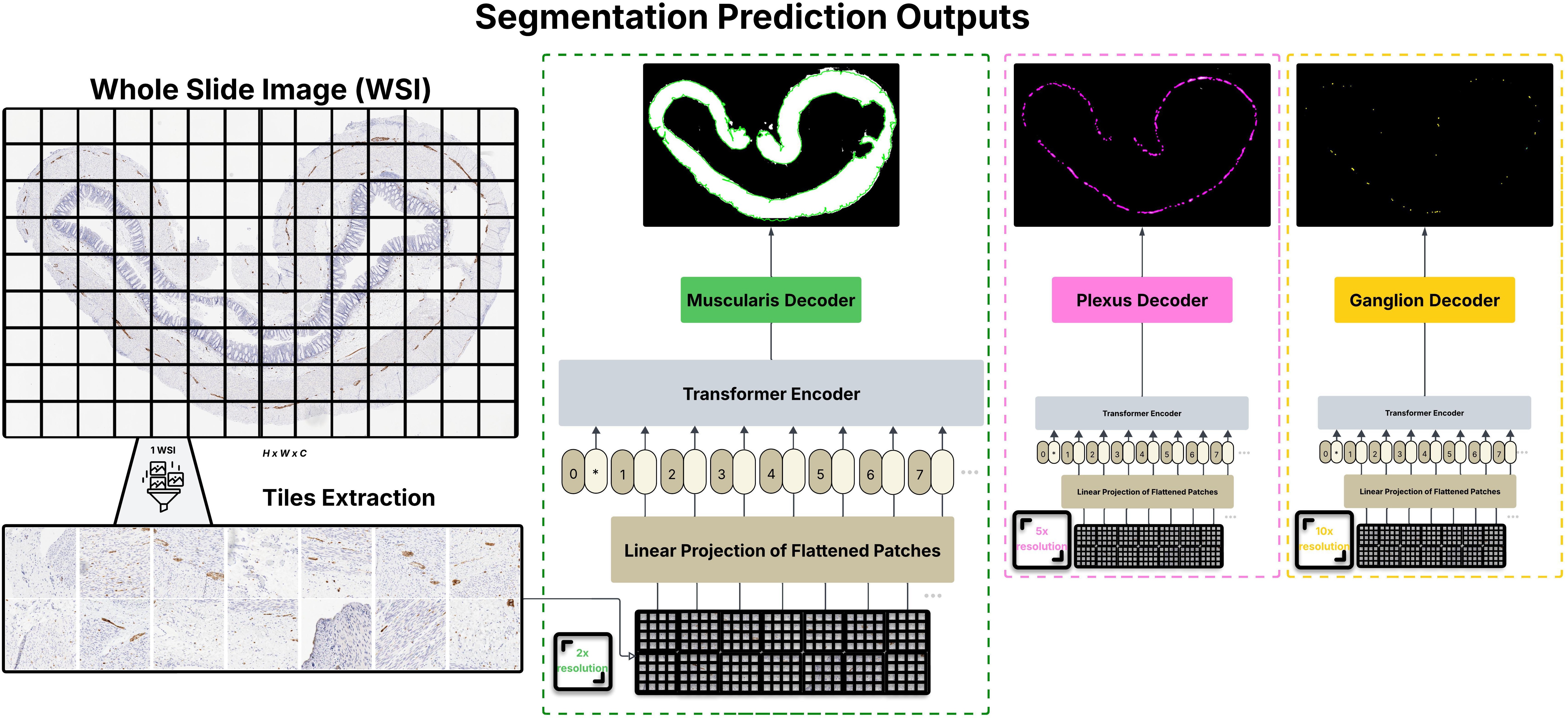}
	\caption{Overview of the model pipeline for WSI processing and multi-target segmentation. The WSI is divided into fixed-size tiles that are linearly projected into patch embeddings before entering the transformer encoder. The encoded representations are passed to three task-specific decoders responsible for muscularis segmentation, plexus segmentation, and ganglion detection. The outputs of these decoders generate the corresponding prediction masks shown at the top.}
	\label{FIG:model_arc}
\end{figure*}

The three-stage analysis framework was based on a pretrained ViT-B/16 backbone{~\hypersetup{hidelinks}\textcolor{blue}{\cite{b17}}}. Specifically, each 224$\times$224 tile was divided into fixed-sized patches (16$\times$16) that were linearly projected to patch embeddings and then fed to the transformer encoder. The encoder at each stage produced a global contextual representation, which was used by a task-specific decoder to generate the corresponding prediction map. Because the three stages operate independently, separate ViT-based models were fine-tuned for muscularis segmentation, plexus segmentation, and ganglion cell detection. At each stage, the decoder received the encoded representations and produced a dense, pixel-wise output for that anatomical layer. An overview of the complete three-stage pipeline is provided in{~\hypersetup{hidelinks}\textcolor{blue}{Fig.~\ref{FIG:model_arc}}}.

\subsubsection{Muscularis Propria}

To ensure robust segmentation of the muscularis propria and to reduce the risk of overfitting given the limited number of WSIs, a 5-fold cross-validation strategy was employed at the WSI level. The dataset was partitioned into five disjoint folds, each containing 6 WSIs. In each iteration, four folds were used for model training, while the remaining fold was held out for evaluation. From each WSI in the training folds, 1000 tiles were randomly extracted and used for model optimization, resulting in 24,000 training tiles per fold. Tiles extracted from the held-out WSIs were not used to update model parameters and were employed only to monitor validation performance during training and to guide early stopping.

Final model evaluation was performed at the WSI level rather than at the tile level. For evaluation, segmentation was carried out using overlapping tiles across each held-out WSI, with a tile size of $224 \times 224$ pixels and a stride of 112 pixels. Each tile was segmented by the ViT model, and only the central $112 \times 112$ pixel region of each predicted tile was retained. These central predictions were then stitched together to reconstruct a full-resolution WSI segmentation map. Quantitative performance metrics were computed by comparing the reconstructed WSI-level prediction maps with the corresponding ground truth segmentation masks. This process was repeated across folds so that each WSI served as the evaluation case exactly once. The evaluation process was used in the other two stages as well (myenteric plexus segmentation and ganglion cells detection).

To increase dataset diversity, the tile training sets were augmented using four methods: scaling, random rotation, horizontal flipping, and vertical flipping. The scaling operation simulated zoom effects by randomly shrinking the visible region by at least 50\% before resizing the tile back to 224$\times$224 pixels. This augmentation enabled the model to learn from tiles representing different effective magnifications. Rotation entailed rotating the tiles by 90$\degree$; horizontal and vertical flipping consisted of flipping the images by 180$\degree$, with a $\frac{1}{2}$ chance of being flipped. Due to the re-stitched prediction masks being binary, with white representing regions of interest and black representing areas not of interest, any rotation operation on the original ground truth mask tiles could result in inverting the areas of interest to be considered as regions not of interest by adding extra black space on the tile borders, thus introducing noise to the true performance. To overcome this, a function was deployed to trim the border regions of the augmented mask, thereby preserving the alignment between the training tile and the ground truth binary mask and reducing potential inconsistent tissue placement. 

For model training, the ViT-B/16 utilized a base learning rate of 5e-4 as the first hyperparameter. An AdamW optimizer was employed, along with a weight decay of 1e-4. Furthermore, a cosine learning rate was used, gradually adjusting the learning rate over time, which converges faster and potentially yields better results. Training used five epochs for warm-up, followed by a complete run of 50 epochs with a batch size of 64. 

Since the muscularis is the largest region within the colon, the tiles that were accepted correspond to random tiles; therefore, the tiles chosen for the model's training purposes weren't necessarily selected based on whether they contained plexus regions. After inputting each tile into the ViT model, the output consists of a pixel-wise raw logit, which is then converted to a probability between 0 and 1 after passing through a SoftMax activation function, representing the probability of the muscularis propria layer. The center region of the prediction tile is then retained and later used for re-stitching to form the full WSI prediction map. To evaluate the model's predictive performance for each layer, a manual sweep operation was performed to ensure proper alignment and no missing regions between the ground truth segmentation binary map and the prediction segmentation binary mask.

\subsubsection{Myenteric Plexus}
Similar to the training procedure for segmenting the muscularis propria, a 5-fold cross-validation strategy was also applied throughout training (24 WSIs, 24,000 tiles) to minimize the risk of the pre-trained ViT model overfitting to the dataset. Identical data augmentation techniques were applied when preparing the tiles for training purposes. This also included scaling, a 90$\degree$ random rotation, and horizontal and vertical flipping of approximately 180$\degree$, each with a 50\% chance of occurrence. To mitigate any potential effects that rotation operation could have had on performance, the border edges of the binary mask were trimmed to preserve alignment between the original tiles and ground truth tiles, specifically in regions of interest. The ViT-based model for segmentation of the myenteric plexus also utilized an AdamW optimizer. Unlike the segmentation of the muscularis propria, segmenting the myenteric plexus utilized a learning rate of 2e-4 and a weight decay of 1e-3. Similarly, a cosine learning rate scheduler was also used with five warmup epochs followed by a full training session of 50 epochs.  

The ViT-based model output comprised pixel-wise logits per tile, which were converted to probabilities (0-1) via a SoftMax activation function, corresponding to the likelihood that a pixel represented the myenteric plexus layer. The tiles are then stitched after preserving their central regions, forming the WSI binary prediction map. A manual sweep is conducted, comparing the model’s prediction to the ground-truth mask to evaluate segmentation performance. As with the muscularis propria layer, a threshold sweep was conducted to determine the optimal pixel value to accept as part of the myenteric plexus layer.

\begin{figure}
	\centering
	\includegraphics[width=0.48\textwidth]{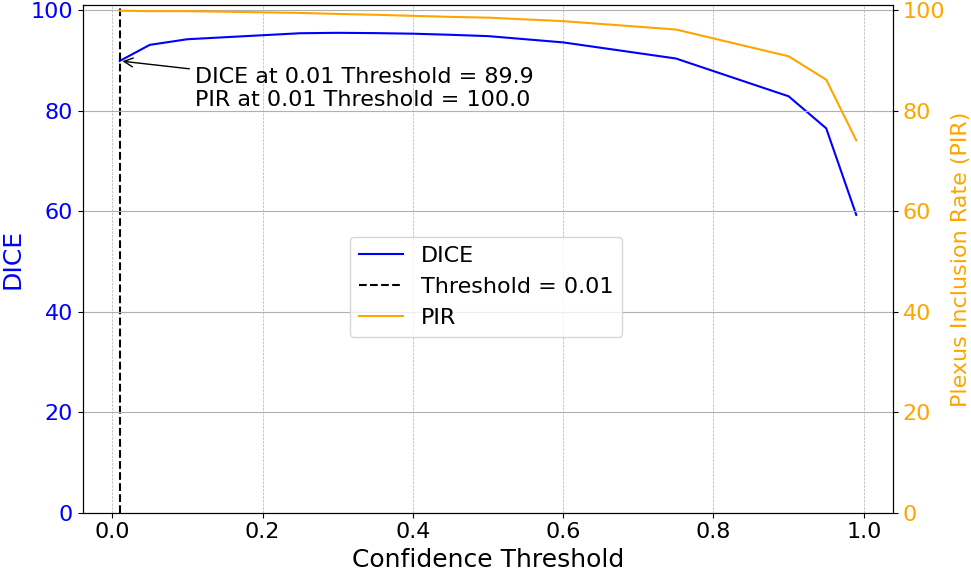}
	\caption{Impact of confidence threshold on Dice coefficient and PIR for the Muscularis Propria stage.}
	\label{FIG:muscle_sweep_figure}
\end{figure}

\begin{figure}
	\centering
	\includegraphics[width=0.48\textwidth]{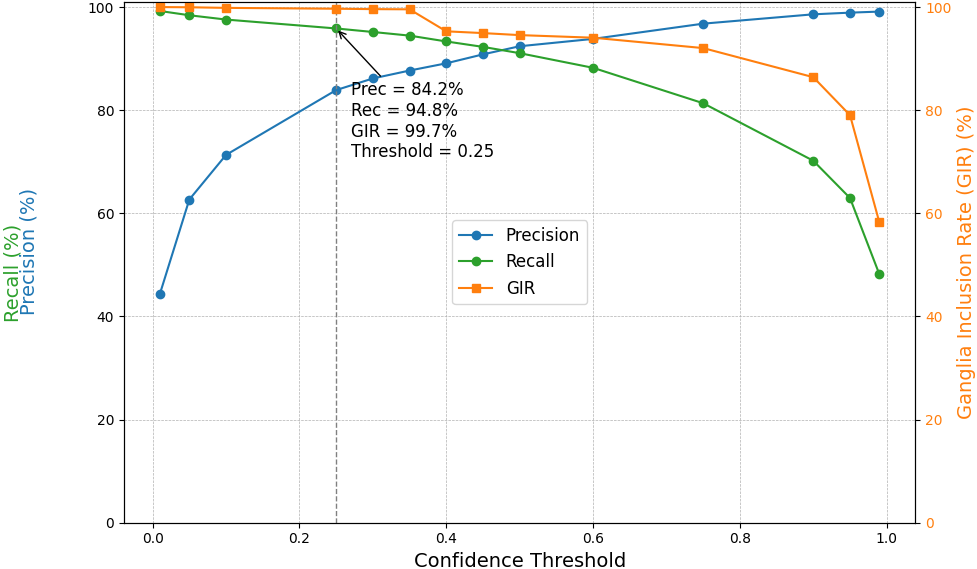}
	\caption{Impact of confidence threshold on Precision and Recall for the Myenteric Plexus stage.}
	\label{FIG:threshold_sweep_plexus_combined}
\end{figure}

\begin{figure}
	\centering
	\includegraphics[width=0.48\textwidth]{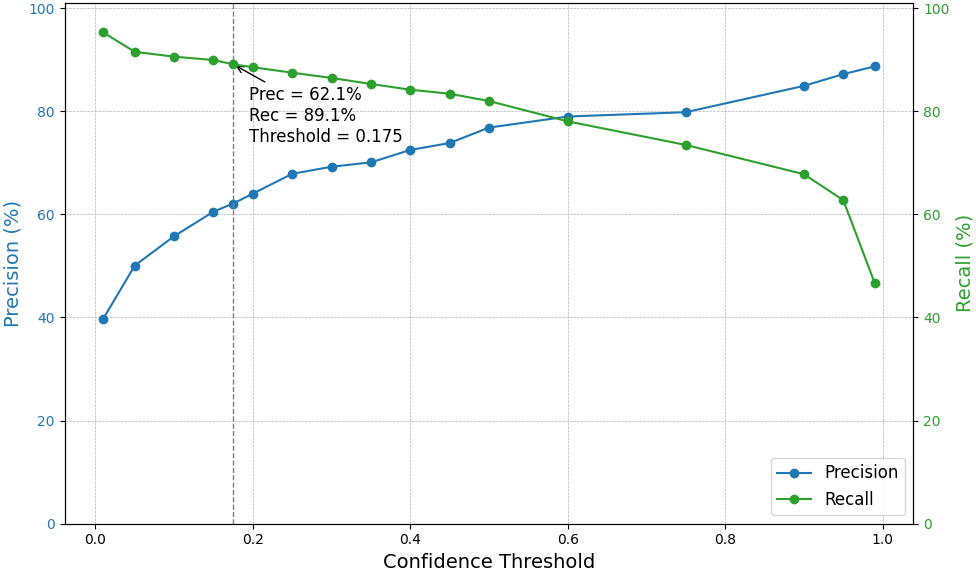}
	\caption{Impact of confidence threshold on Precision and Recall for the Ganglion Cells stage.}
	\label{FIG:threshold_sweep_ganglia}
\end{figure}

\subsubsection{Ganglion Cells}
Similar to the muscularis propria and plexus segmentation tasks, ganglion cell detection was performed using a 5-fold cross-validation strategy at the WSI level. In each fold, 1000 tiles were randomly extracted from each training WSI and used for model optimization, while tiles from the held-out WSIs were reserved for validation and WSI-level evaluation. Final performance was assessed by re-stitching overlapping tile predictions to form full-resolution WSI segmentation maps.

To increase data diversity, the 224$\times$224-pixel tiles were augmented in four different ways. The first augmentation technique was scaling, followed by vertical and horizontal flipping by 180$\degree$, and rotation, which rotated an image by 180$\degree$ with a $\frac{1}{2}$ chance. Since the augmentations were also applied to binary masks, border trimming was performed to prevent flipping operations from inverting regions of interest and creating inconsistencies. Hyperparameters included a learning rate of 5e-4, a weight decay of 1e-4, and the AdamW optimizer. Training consisted of 5 warm-up epochs followed by 50 full training epochs, with a batch size of 64.

The ViT-based model output consisted of raw pixel logits, which were converted to probabilities between 0 and 1 via a SoftMax activation. The center regions of the tiles were preserved and re-stitched to reconstruct the full prediction mask. Tiles were then accepted as ganglion cell predictions if their probabilities exceeded the predetermined threshold. To evaluate predictive performance, a manual sweep was conducted to compare the stitched prediction maps with the ground truth segmentation maps.

\subsection{Data Postprocessing}
\subsubsection{Myenteric Plexus}
Binary mask tiles produced by the ViT model were first processed by retaining their center regions and re-stitching them to reconstruct the full WSI prediction map. Connected component labelling was applied to group adjacent positive pixels into individual plexus regions, and any predicted component with an area smaller than 50 pixels was removed to eliminate small spurious detections. This object-wise filtering step ensured that only anatomically plausible plexus regions were retained in the final prediction mask.

\subsubsection{Ganglion Cells}
Postprocessing for ganglion cell prediction followed the same initial procedure of retaining tile centers and re-stitching them to reconstruct the full WSI-level mask. Predicted ganglion regions were restricted to the previously segmented myenteric plexus by applying a binary mask, ensuring that detections were only evaluated within anatomically valid regions. Connected component labeling was then used to group positive pixels into individual ganglion candidates. To suppress small spurious detections, any predicted component with an area smaller than 10 pixels was removed. This object-level filtering step yielded a cleaner, anatomically consistent set of ganglion predictions without altering the morphology of the retained objects.

\subsection{Performance Metrics}
To evaluate segmentation performance across all three layers, several common metrics were used. Precision and recall quantify the model’s ability to correctly identify positive pixels, defined as
\begin{equation}
\text{Precision} = \frac{TP}{TP + FP}
\end{equation}
\begin{equation}
\text{Recall} = \frac{TP}{TP + FN}
\end{equation}
where a True Positive (TP) corresponds to a correctly identified positive pixel, a False Positive (FP) corresponds to a pixel incorrectly labeled as positive, and a False Negative (FN) represents a pixel that should have been labeled positive but was not.

\begin{figure*}
	\centering
	\includegraphics[width=0.8\textwidth]{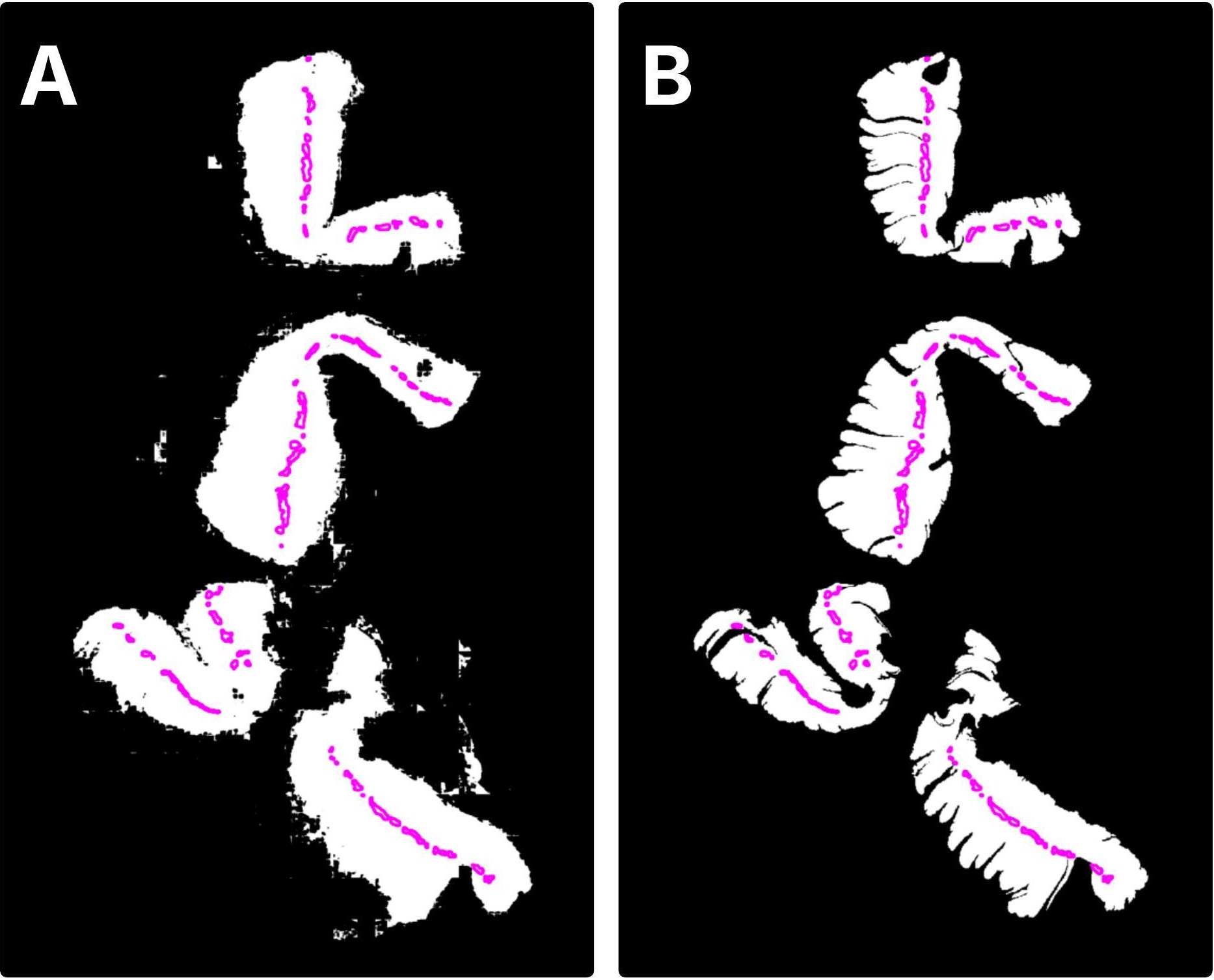}
	\caption{Muscularis propria layer segmentation on a whole slide section with plexus annotations overlaid.
    \textbf{A:} Stitched model prediction for the muscularis layer shown in white, with ground truth plexus regions overlaid in magenta.
    \textbf{B:} Ground truth muscularis mask with the same plexus regions overlaid in magenta. 
    The magenta plexus annotations appear consistently in both panels, reflecting that these regions are fully captured by the model, yielding a perfect PIR.}
	\label{FIG:muscle_pred_example}
\end{figure*}

Layer-specific metrics were also used where appropriate. The Dice coefficient measures spatial overlap:
\begin{equation}
\mathrm{Dice} = \frac{2|X_{ViT} \cap Y_{GT}|}{|X_{ViT}| + |Y_{GT}|}
\end{equation}
where $X_{ViT}$ is the predicted mask and $Y_{GT}$ is the ground truth mask.

For evaluating the inclusion of relevant anatomical regions, two additional metrics were used.
The Plexus Inclusion Rate (PIR) for muscularis segmentation is defined as
\begin{equation}
\mathrm{PIR} = \frac{N_{ViT}}{N_{GT}},
\end{equation}
where $N_{GT}$ is the total number of ground truth plexus regions and $N_{ViT}$ is the number of those plexus regions that intersect the predicted muscularis mask at least once.
Similarly, the Ganglia Inclusion Rate (GIR) for plexus segmentation is defined as
\begin{equation}
\mathrm{GIR} = \frac{N_{ViT}}{N_{GT}},
\end{equation}
where $N_{GT}$ is the total number of ground truth ganglion regions and $N_{ViT}$ is the number of those ganglion regions that are fully contained within, or intersect, the predicted plexus mask.

\subsubsection{Muscularis Propria}
For muscularis segmentation, performance was assessed using the Dice coefficient, PIR, precision, and recall. The Dice coefficient measured the overlap between the predicted and ground-truth muscularis regions. At the same time, PIR quantified how effectively the segmented muscularis retained the plexus regions, which is necessary for downstream layers. Precision and recall quantified pixel-level accuracy.

\subsubsection{Myenteric Plexus}
For myenteric plexus segmentation, precision, recall, and GIR were used. Precision measured how many predicted plexus pixels were correct, while recall measured how completely the plexus regions were captured. GIR quantified how many ganglion-containing plexus regions were retained, which directly affects the success of ganglion detection.

\subsubsection{Ganglion Cells}
Object-level accuracy was calculated for ganglion cell detection using precision and recall, defined in Equations (1) and (2). In this case, a TP is a predicted ganglion object that has an overlap with a ground truth ganglion annotation. A FP is a predicted ganglion object that has no overlap with any ground truth ganglion, and a FN is a ground truth ganglion that was not detected by the model. Precision, then, can be interpreted as how many of the predicted ganglion objects are correct, while recall indicates how completely the model is capturing the annotated ganglion cells.

\subsection{Baseline Method}
To contextualize the performance of the proposed ViT-based pipeline, results were compared against previously published methods that evaluated the same WSI dataset {\hypersetup{hidelinks}\textcolor{blue}{\cite{b11,b13,b14}}}. These baseline approaches are summarized below.

\subsubsection{Muscularis Propria}
The baseline model's methodology to successfully segment the muscularis propria consisted of utilizing Calretinin-stained WSI of the colon cross-sections and deploying a colour-based k-means clustering algorithm to segment the colon section{~\hypersetup{hidelinks}\textcolor{blue}{\cite{b14}}}. Additionally, a CNN-based algorithm that was pretrained on ImageNet-1k was also utilized to segment the muscularis propria layer {\hypersetup{hidelinks}\textcolor{blue}{\cite{b13,b14}}}, using tiles with dimensions of 256$\times$256 pixels. Postprocessing techniques were also applied, which included morphological smoothing{~\hypersetup{hidelinks}\textcolor{blue}{\cite{b19}}}.

\begin{table*}[htbp]
\centering
\caption{Performance comparison (\%) of segmentation models on the \textbf{Muscularis Propria} layer.}
\label{tab:muscularis_propria_results}
\setlength{\tabcolsep}{10pt}
\renewcommand{\arraystretch}{1.5}
\begin{tabular}{lcccccc}
\toprule

\textbf{Model} & \textbf{Dice} & \textbf{Precision} & \textbf{Recall} & \textbf{PIR} \\
\midrule

K-means{~\hypersetup{hidelinks}\textcolor{blue}{\cite{b14}}} & 70.7 & 70.6 & 78.9 & 77.4 \\
CNN{~\hypersetup{hidelinks}\textcolor{blue}{\cite{b13,b14}}} & 89.2 & 81.9 & 96.2 & 96.0 \\
\textbf{ViT-B/16 (proposed)}{~\hypersetup{hidelinks}\textcolor{blue}{\cite{b15}}} & \textbf{89.9} & \textbf{82.4} & \textbf{99.7} & \textbf{100} \\

\bottomrule
\end{tabular}
\end{table*}

\subsubsection{Myenteric Plexus}
The segmentation of the myenteric plexus involved receiving the binary masks, corresponding to the model's prediction output, and utilizing these images for the segmentation process. Unlike the muscularis propria segmentation, the myenteric plexus was segmented first, using a colour-based thresholding to differentiate between brown-stained areas{~\hypersetup{hidelinks}\textcolor{blue}{\cite{b14}}}. This was then followed by applying a pixel threshold to accept and reject pixels that could correspond to being a part of the myenteric plexus layer. Furthermore, morphological filtering{~\hypersetup{hidelinks}\textcolor{blue}{\cite{b19}}} was also applied to filter out pixels that potentially correspond to true or false myenteric plexus regions for the final stage of segmentation.

\subsubsection{Ganglion Cells}
The final step included classifying true ganglion cells within each plexus region, also utilizing a colour intensity-based threshold to determine candidates for ganglion cells {\hypersetup{hidelinks}\textcolor{blue}{\cite{b14}}}, fitting categories with different certainties, high-certainty is represented by being compact, with a visible nuclei. Whereas, low-certainty candidates are borderline ganglion, which do appear to be ganglion; however, they do lack typical morphology and ganglion physical traits. Furthermore, physical characteristics such as area, circularity, colour, and gradient contrast were used as descriptors for each candidate, and then used to train an LDA model to classify true ganglion cells from false candidates.

\section{Results}
This research study’s results section aims to evaluate the segmentation performance of the ViT-B/16 model across three layers of the colon: muscularis propria, myenteric plexus, and ganglion cells. The metrics presented pertain to each segmented layer. 

\subsection{Muscularis Propria}
The segmentation of the muscularis propria region is of significant importance, as it shapes the anatomical boundaries for the later identification of ganglion cells located within the myenteric plexus. Therefore, isolating the muscularis layer allows segmentation of the plexus region, where ganglion cells are typically found. Furthermore, this narrows the region of interest from a much broader WSI to only tissue regions clinically more relevant to diagnosing HD, as evidenced by a high Dice coefficient. Moreover, confidence thresholds, which determine the model’s ability to include a tile as part of the muscularis propria, were experimented with. In terms of determining the most optimal threshold to accept pixels corresponding to a muscularis propria, a threshold sweep of 0.01-0.99 was applied, and the threshold which produced the highest PIR and Dice coefficient was utilized. The effect of varying the prediction threshold on both Dice and PIR is summarized in{~\hypersetup{hidelinks}\textcolor{blue}{Fig.~\ref{FIG:muscle_sweep_figure}}}. A confidence threshold of 0.4 yields model results that are similar to those of the ground truth. However, a lower confidence threshold of 0.01 achieves a superior PIR with a lower Dice coefficient (chosen as the optimal value), giving further certainty that all tiles include a plexus region, which serves the ultimate goal of detecting ganglion cells better, as shown in{~\hypersetup{hidelinks}\textcolor{blue}{Fig.~\ref{FIG:muscle_sweep_figure}}}. 

The results, as shown in {\hypersetup{hidelinks}\textcolor{blue}{Table~\ref{tab:muscularis_propria_results}}}, demonstrate a high Dice coefficient of 89.9\%, indicating a more accurate segmentation performance of the muscularis layer compared to the ground truth manual segmentation. This not only shows a high overall accuracy but also superior performance compared to the baseline CNN and K-means models. Additionally, the model achieved a precision score of 82.4\%, indicating that the muscularis region corresponds to muscularis tissue, mislabeling only a few non-muscularis tissues. The model also achieved a recall score of 99.7\%, demonstrating its superior ability to identify all positive tissue pixels. The model also achieved a PIR of 100\%. That said, all model-predicted outputs contained plexus regions, which play an instrumental role in the subsequent segmentation stage. A qualitative comparison of predicted muscularis segmentation with plexus overlays is shown in{~\hypersetup{hidelinks}\textcolor{blue}{Fig.~\ref{FIG:muscle_pred_example}}}.

\begin{table}[htbp]
\centering
\caption{Performance comparison (\%) of segmentation performance on \textbf{Myenteric Plexus} layer.}
\label{tab:myenteric_plexus_results}
\setlength{\tabcolsep}{4pt}
\renewcommand{\arraystretch}{1.5}
\begin{tabular}{lcccccc}
\toprule

\textbf{Model} & \textbf{Precision} & \textbf{Recall} & \textbf{GIR} \\
\midrule

Colour-based thresholding {\hypersetup{hidelinks}\textcolor{blue}{\cite{b11,b14}}} & 73.8 & 85.9 & 99.2 \\
\textbf{ViT-B/16 (proposed)} & \textbf{84.2} & \textbf{94.8} & \textbf{99.7} \\

\bottomrule
\end{tabular}
\end{table}

\subsection{Myenteric Plexus}
\begin{figure*}
	\centering
	\includegraphics[width=0.98\textwidth]{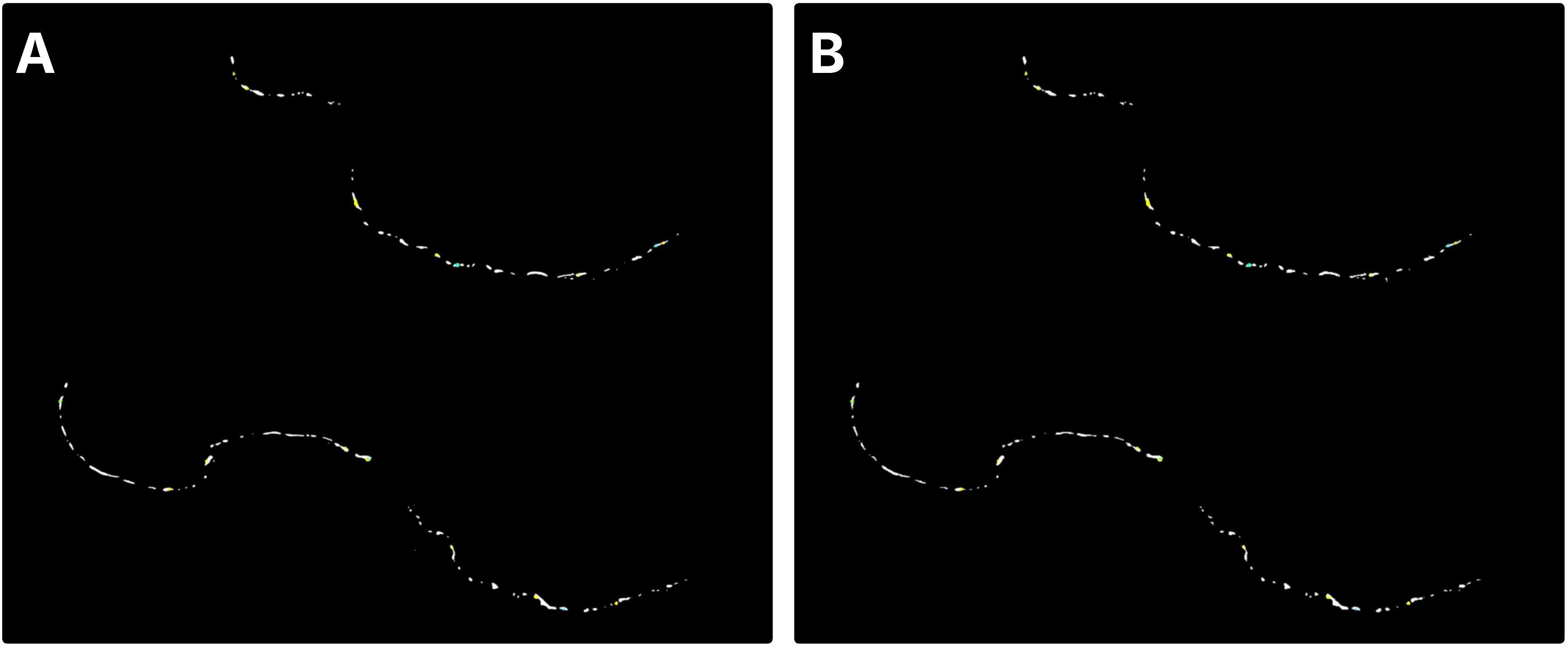}
	\caption{Myenteric plexus region segmentation on a whole slide section with ganglion cell annotations overlaid.
    \textbf{A:} Stitched model prediction for the plexus regions displayed in white.
    \textbf{B:} Corresponding ground truth mask for the plexus regions.
    Yellow and cyan objects represent high- and low-certainty ganglion cells, respectively. All 28 ganglion cells fall within the predicted plexus regions, demonstrating complete ganglion inclusion in the plexus segmentation.}
	\label{FIG:plexus_example_pred}
\end{figure*}

The following segmentation stage included segmenting the myenteric plexus section of the colon. This stage is considered the most crucial segmentation phase before ganglion identification, as ganglion cells are located within the myenteric plexus layer. The ViT-B/16 proposed model achieves high segmentation accuracy in the plexus layer, which prepares for optimal ganglia identification in the downstream task. A confidence threshold of 0.25 was used to select the predicted pixels in the binary output prediction map. Although the chosen confidence threshold was flexible, it yielded a higher recall at the expense of precision, prioritizing a higher GIR score ({\hypersetup{hidelinks}\textcolor{blue}{Fig.~\ref{FIG:threshold_sweep_plexus_combined}}}). This threshold yielded the optimal GIR, as well as precision and recall. A representative plexus prediction example with ganglion cell annotations is shown in{~\hypersetup{hidelinks}\textcolor{blue}{Fig.~\ref{FIG:plexus_example_pred}}}.

The results, as shown in {\hypersetup{hidelinks}\textcolor{blue}{Table~\ref{tab:myenteric_plexus_results}}}, demonstrate a high recall score of 94.8\%, indicating the model’s ability to detect almost all positive samples, corresponding to regions that contain myenteric plexus. Furthermore, a high precision score of 84.2\% demonstrates the model’s capability of detecting regions that not only accurately pertain to the myenteric plexus but also limit false positives by being confined to regions enclosed by the muscularis. Finally, the ViT model achieved a high GIR score of 99.7\% when segmenting the myenteric plexus. Given the myenteric plexus’ thin and long anatomical structure, and its thin, elongated, and discontinuous nature, the model’s segmentation performance, with high GIR ({~\hypersetup{hidelinks}\textcolor{blue}{Fig.~\ref{FIG:plexus_example_pred}}}), precision, and recall, allows for higher chances of success in later ganglion cell identification. 

\subsection{Ganglion Cells}
\begin{figure*}
	\centering
	\includegraphics[width=0.8\textwidth]{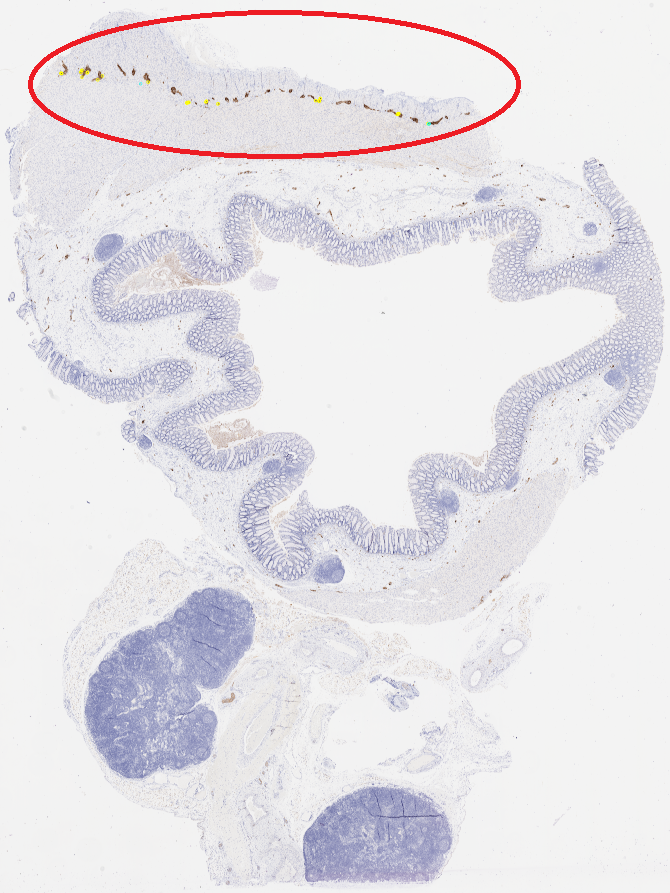}
	\caption{WSI with predicted ganglion cells overlaid. Cyan objects represent low certainty predictions, and yellow objects represent high certainty predictions. All visible ganglion cells in this section are correctly identified by the model.}
	\label{FIG:ganglia_pred_example}
\end{figure*}

Once the myenteric plexus was segmented, the next stage involved segmenting and identifying ganglion cells to further enhance clinical relevance to HD. This segmentation stage involved transitioning from region-level segmentation to cell-level (object) prediction. The threshold utilized for segmenting and detecting the ganglion cells is 0.175 ({\hypersetup{hidelinks}\textcolor{blue}{Fig.~\ref{FIG:threshold_sweep_ganglia}}}). An example visualization of predicted high- and low-certainty ganglion cells across a whole slide is shown in{~\hypersetup{hidelinks}\textcolor{blue}{Fig.~\ref{FIG:ganglia_pred_example}}}.

\begin{table*}[htbp]
\centering
\caption{Performance comparison (\%) of segmentation performance on \textbf{Ganglion Cell} identification.}
\label{tab:ganglion_cell_results}
\setlength{\tabcolsep}{6pt}
\renewcommand{\arraystretch}{1.5}
\begin{tabular}{lcccccc}
\toprule

\textbf{Model} & \textbf{Precision (high-certainty)} & \textbf{Recall (high-certainty)} & \textbf{Precision (combined)} & \textbf{Recall (combined)} \\
\midrule

LDA model {\hypersetup{hidelinks}\textcolor{blue}{\cite{b14}}} & 60.9 & 82.1 & 64.8 & \textbf{80.2} \\
\textbf{ViT-B/16 (proposed)} & \textbf{62.1} & \textbf{89.1} & \textbf{67.0} & 78.6 \\

\bottomrule
\end{tabular}
\end{table*}


The ViT-B/16 proposed model achieved a high recall average score of 78.6\% ({\hypersetup{hidelinks}\textcolor{blue}{Table~\ref{tab:ganglion_cell_results}}}) by utilizing the combined low-certainty high-certainty ganglion cells, demonstrating strong capabilities in detecting the majority of ganglion cells within the myenteric plexus region. Furthermore, using a more conservative threshold for higher certainty yields a precision of 62.1\%, indicating the model’s ability to identify clear, morphologically distinct ganglion cells. Using the same high-certainty threshold, the model achieved a high recall of 89.1\%, thereby identifying the overwhelming majority of ganglion cells with a typical ganglion-cell morphology. 
Overall, the DL model used for segmenting and identifying ganglion cells outperformed the baseline model in ganglion cell detection.

\section{Discussion}
\subsection{Interpretation of Results}
This research study proposed a multi-stage pipeline to segment three colon layers, with the overall goal of identifying the presence or absence of ganglion cells. The segmentation process used a ViT-based model to first segment the muscularis propria, then the myenteric plexus, and finally the ganglion cells within the myenteric plexus. The proposed approach achieved high Dice coefficient and PIR scores when segmenting the muscularis, a high GIR score when segmenting the myenteric plexus, and, finally, high precision and recall scores for identifying ganglion cells within the plexus region. Demonstrating the model’s capability to detect fine colon boundaries and detect ganglia with confidence. Performance evaluations surpassed those of conventional CNN methods and other shallow machine learning methods, indicating superior performance for segmenting colon tissue. These findings further demonstrate the potential of DL algorithms to aid HD diagnosis. 

The muscularis propria is a smooth muscle layer embedded within the walls of organs such as the GI tract and bladder, composed of an inner circular layer and an outer longitudinal layer, giving it a more uniform structure {\hypersetup{hidelinks}\textcolor{blue}{\cite{b26}}}. The ViT has an inherent self-attention mechanism, enabling it to attend to other image patches and capture long-range dependencies. For instance, it can relate the appearance of the muscularis propria to its spatial context relative to adjacent layers, such as the lamina propria or submucosa. The segmentation performance of the ViT model yielded a Dice coefficient of 89.9\%, indicating strong agreement between the predicted and ground truth maps. Notably, the model slightly oversegments the muscularis, reflected in higher recall and lower precision. This behaviour increases the number of false positive muscularis pixels, but it preserves a perfect PIR, ensuring that all plexus regions remain fully contained within the predicted muscularis mask. While this enlarges the search region for stage two, the tradeoff is acceptable, as it prioritizes anatomical completeness needed for downstream segmentation. These results demonstrate that global-context information, combined with local feature representations, enables ViTs to achieve high segmentation accuracy, as evidenced by the Dice coefficients obtained. Additionally, segmentation performance on this layer yields a PIR of 100\%, ensuring that the plexus segmentation is applied within the anatomical region of interest for later accurate segmentation of ganglion cells. 

The myenteric plexus region is embedded within the inner layer of the muscularis propria and consists of nerve fibres responsible for several autonomous digestive functions. Unlike the muscularis propria, the myenteric plexus is a smaller, less structured region and thus appears more fragmented across WSIs {\hypersetup{hidelinks}\textcolor{blue}{\cite{b26,b27}}}, making segmentation more difficult. The myenteric plexus is of great importance because it is the anatomical site where ganglion cells are present {\hypersetup{hidelinks}\textcolor{blue}{\cite{b27}}}, which are needed for future segmentation and identification. The segmentation performance yielded recall and precision scores of 94.8\% and 84.2\%, respectively. Indicating the model’s robust capability of segmenting most of the pixels that correspond to the plexus region. 
Additionally, among the tiles identified as plexus, the model produces relatively few false positive segmentations. When oversegmentation does occur, where the predicted plexus region extends slightly beyond the true anatomical boundary, this is generally acceptable because it still maintains a high GIR. In this setting, including a larger area around each plexus reduces the likelihood that a plexus containing ganglion cells will be clipped or partially excluded. This behaviour tends to lower precision, since more pixels are labelled as plexus, but it preserves the downstream objective of ensuring that ganglion-containing regions are retained.

The recall observed (94.8\%) for plexus likely reflects instances where entire plexus regions were missed rather than small local errors. This can be examined by recomputing recall after removing completely missed plexus objects from the evaluation, thereby yielding a higher recall value. While some plexus regions are not detected, the very high GIR of 99.7\% indicates that the missed regions do not contain ganglion cells and therefore do not influence the success of stage three. It is also possible that the ground truth plexus annotations in these cases were conservative, capturing regions that are anatomically adjacent to the plexus but not functionally relevant. As a result, the model may be omitting less important or marginal areas without affecting the overall goal of preserving ganglion-containing plexus structures. Because the myenteric plexus is embedded within the muscularis propria, producing textural variability {\hypersetup{hidelinks}\textcolor{blue}{\cite{b26}}}, this allows ViTs to leverage their inherent self-attention mechanisms, achieving superior results. For instance, the attention layers embedded within the ViT enable the model to distinguish between the thick, continuous, and homogeneous fibres of the muscularis layer and the thin, irregular fibres of the plexus, even when the muscularis propria and the myenteric plexus share the exact colour tones {\hypersetup{hidelinks}\textcolor{blue}{\cite{b28}}}. The ViT computes attention scores between patch pairs, enabling it to identify which patches are similar or different based on the embeddings it produces. This allows the model to distinguish between slight variations in texture.
Furthermore, given the fragmented myenteric plexus throughout the WSI, it is discontinuous across several patches, providing ViTs with an opportunity to leverage their inherent understanding of global contextual relationships. Unlike conventional CNNs, which primarily rely on local receptive fields, this may prevent recognizing that spatially separated patches belong to the same structure, leading to fragmented predictions, especially when there are mild structural differences between patches. However, since ViTs operate globally, the model can label patches with slightly different structures as belonging to the plexus region, even when their embeddings are similar. That being said, high segmentation accuracy and high GIR of the myenteric plexus are critical to optimizing ganglion cell segmentation and identification. 

The final stage of segmentation is considered the most biologically significant, aiming to identify the presence or absence of ganglion cells, which is directly attributed to the clinical diagnosis of HD. Unlike the two previous stages of segmentation, ganglion cell segmentation transitions from region-level segmentation to object-level segmentation of cells. Due to morphological differences between different ganglion cells, such as their shape, size and sparse distribution within the myenteric plexus {\hypersetup{hidelinks}\textcolor{blue}{\cite{b29}}}, their segmentation is increasingly complex compared to previous layers. Segmenting the ganglion cells used two different manual annotations: one for ganglion cells classified as "high-certainty", thus possessing typical ganglion morphological characteristics. Whilst "low-certainty" is utilized for ganglion cells that lack typical morphological representations. At a high confidence level, the model achieves 62.1\% precision and 89.1\% recall. The average of the high- and low-certainty results yields recall and precision scores of 78.6\% and 67.0\%, respectively. Based on these results, the model is more conservative, producing fewer but highly reliable detections.
On the other hand, a less strict threshold ensures that most ganglion cells are predicted, at the cost of limited reliability. Unlike previous layers for segmentation, where the ViT model leveraged its innate self-attention mechanism to distinguish local regions, for this segmentation task, the ViT leverages its innate mechanism to understand morphological cues corresponding to ganglion cell characteristics by integrating fine-grained texture into feature embeddings. For instance, the self-attention mechanism for this task interprets patterns of nuclei, dendrites, and axons, thus focusing on local micro-relationships rather than long-range dependencies. Furthermore, the ViT architecture implements certainty thresholds for ganglion cells, where the embedded attention heads agree on whether the prediction is most likely a ganglion cell; in low-certainty mode, the attention heads produce more dispersed attention on the segmented prediction. This two-tier framework shows how these results, in terms of certainty, might be helpful for diagnostic confirmation or further review, similar to applications in a clinical setting. Ganglion cell identification using DL demonstrates the potential to automate the most critical step in diagnosing HD, reducing diagnostic variability and further assisting pathologists. 

Previous research has used similar approaches to segment and detect the potential presence of ganglion cells {\hypersetup{hidelinks}\textcolor{blue}{\cite{b13,b14}}}. For instance, a traditional CNN model {\hypersetup{hidelinks}\textcolor{blue}{\cite{b13,b14}}} was utilized to segment the muscularis propria layer alongside a shallow learning model, a K-means clustering algorithm {\hypersetup{hidelinks}\textcolor{blue}{\cite{b14}}}. This resulted in the CNN model producing a Dice coefficient of 89.2\%, a precision of 81.9\%, a recall of 96.2\% and a PIR of 96\%. Compared to our 89.9\%, 82.4\%, 99.7\% and 100\%, respectively. Performance regarding the myenteric plexus tissue in {\hypersetup{hidelinks}\textcolor{blue}{\cite{b11,b14}}}, utilized a colour-based threshold framework and produced a recall score of 85.9\%, precision of 73.8\%, GIR of 99.2\%. Compared to our 94.8\%, 84.2\%, and 99.7\%, respectively. Regarding the ganglion segmentation, for the LDA model {\hypersetup{hidelinks}\textcolor{blue}{\cite{b14}}}, the combined certainty ganglia (high and low certainty) produced a recall of 80.2\% and precision of 64.8\%, compared to our 78.6\% and 67.0\%, respectively. For the high-certainty ganglion segmentation results reported in {\hypersetup{hidelinks}\textcolor{blue}{\cite{b14}}}, precision was 60.9\% and recall was 82.1\%, compared to our 62.1\% and 89.1\%. Results from this study highlight the ViT's ability to leverage its inherent self-attention mechanism to learn global dependencies across an image. This allows the model to accurately isolate the muscularis propria layer, consistently linking relationships within the fragmented myenteric plexus and understanding local microcellular structure at the microscale. In contrast to conventional CNNs or other shallow learning models, which typically rely on localized features within each map, ViT's architecture enables global reasoning across different structures and textures within the same image or region, thereby achieving higher performance than traditional CNNs for segmentation tasks. 

This research study proposed a multi-stage hierarchical segmentation pipeline. First, segment the muscularis propria layer, then the myenteric plexus layer, and finally segment and identify the presence of ganglion cells within the plexus layer, mirroring the workflow of a pathologist when detecting ganglion cells in the plexus region. Further, the consistent results highlight the potential of DL being utilized in a clinical setting as a confirmatory tool for diagnosing HD. The advantages of this study could be reduced time to interpret WSIs and an automated solution. Reducing inter-observer variability also provides a uniform way to segment and confirm the presence of HD. 

\subsection{Study Limitations}
Although this study demonstrated strong segmentation performance and is deemed promising for future steps, it does include certain limitations that could limit its generalizability to future unseen data. For instance, the data used for this research study were obtained solely from CHEO, as mentioned in Section 2.1, thereby limiting variability in structural and staining differences across WSIs that would have been acquired from different datasets and potentially increasing model generalizability. Furthermore, this study uses a small dataset of 30 WSIs, which increases the potential for overfitting during training, especially since large DL models such as CNNs and ViTs typically yield consistent improvements on larger datasets. Furthermore, since a single pathologist labeled the ground truth masks, this could introduce annotation bias; that said, a variety of expert manual annotations could improve model generalization in future work. Addressing these limitations in future studies through larger multi-center datasets with a variety of staining techniques and structures, a range of expert annotations, and increased dataset size could further enhance model robustness and clinical relevance. 

\subsection{Future Work} 
Although the limitations mentioned above limit the model's generalization ability, the ViT-based model used for segmentation and ganglion cell detection nevertheless produced highly accurate segmentations compared to conventional LDA and other shallow learning models. That said, future research work could benefit from optimizing the utility of ViTs' performance to increase its clinical applicability by incorporating data corresponding to similar forms of colon disease, such as megacystis microcolon intestinal hypoperistalsis syndrome and internal anal sphincter achalasia, giving the model opportunities to learn from different colonic structures and anatomical locations. Furthermore, future models of ViTs could also utilize other forms of self-supervised learning, such as distillation with no labels (DINO I \& II) {\hypersetup{hidelinks}\textcolor{blue}{\cite{b20,b21}}} or other forms of Masked Autoencoding {\hypersetup{hidelinks}\textcolor{blue}{\cite{b22}}}, such as a curriculum masking schedule {\hypersetup{hidelinks}\textcolor{blue}{\cite{b23}}}, both allowing the model to learn essential features corresponding to the structure, texture and location of the colon from solving complex pretext tasks. Collectively, these directions will advance the integration of computer vision into digital pathology, enabling reliable, objective utility for HD diagnosis.

\section{Conclusion}
This study presents a hierarchical segmentation framework that employs a ViT-B/16 model to segment the muscularis propria and myenteric plexus, enabling subsequent detection and identification of ganglion cells, which satisfies a key diagnostic criterion for HD. The workflow mirrors the approach used by pathologists, sequentially isolating the muscularis layer, delineating the plexus regions, and identifying ganglion cells within them. The framework achieved a Dice coefficient of 89.9\% and a PIR of 100\% for muscularis segmentation, a GIR of 99.7\% for plexus segmentation, and a precision and recall of 62.1\% and 89.1\% for high-certainty ganglion cell identification. These findings illustrate ViTs' ability to use self-attention to capture relevant features and global contextual relationships, enabling the model to distinguish fine textural details at both regional and cellular scales. The results further suggest that such a framework could reduce assessment time and diagnostic variability. Although future work would benefit from larger multi-center datasets and multiple expert annotations, this study demonstrates the promise of DL based methods for supporting histological assessment in HD.

\section*{Declaration of competing interest}
The authors declare that they have no known competing financial interests or personal relationships that could have appeared to influence the work reported in this paper.


\end{document}